\documentclass[twocolumn,prb,showpacs,preprintnumbers,amsmath,amssymb]{revtex4}
\usepackage{graphicx}
\usepackage{dcolumn}
\usepackage{bm}
\begin{document}
\title{Tomonaga-Luttinger liquid parameters of magnetic waveguides in graphene}
\author{W. H\"ausler,$^{1,2}$ A. De Martino,$^{1,3}$ T.K. Ghosh,$^{1,4}$ and R. Egger$^1$ }
\affiliation{
$^{1}$Institut f\"ur Theoretische Physik, Heinrich-Heine-Universit\"at, D-40225 D\"usseldorf, Germany \\
$^{2}$Physikalisches Institut, Albert-Ludwigs-Universit\"at, D-79104 Freiburg, Germany\\
$^{3}$Institut f\"ur Theoretische Physik, Universit\"at zu K\"oln, Z\"ulpicher Str.~77, D-50937 K\"oln, Germany \\
$^4$Department of Physics, Indian Institute of Technology-Kanpur, Kanpur 208016, India} 

\date{\today}
\begin{abstract}
Electronic waveguides in graphene formed by counterpropagating
snake states in suitable inhomogeneous magnetic fields are shown
to constitute a realization of a Tomonaga-Luttinger liquid.
Due to the spatial separation of the right- and left-moving
snake states, this non-Fermi liquid state induced by
electron-electron interactions is essentially unaffected by
disorder. We calculate the interaction parameters 
accounting for the absence of Galilei invariance in this system, and 
thereby demonstrate that non-Fermi liquid effects are
significant and tunable in realistic geometries.
\end{abstract}
\maketitle

\section{Introduction}

One-dimensional (1D) electron systems can nowadays be
studied in different material systems, e.g.\
by depositing negatively charged metallic
gate electrodes on top of a 2D electron gas (2DEG)
in semiconducting heterostructures, thereby depleting
the 2DEG to form the desired structure,\cite{ferry05} or in
single-wall nanotubes (SWNTs).\cite{tubes}
Such 1D quantum wires have been argued to 
realize the non-Fermi liquid behavior of a 
{\sl Tomonaga-Luttinger liquid (TLL)},\cite{solyom,haldane,voit,gogolin,giamarchi}
arising as a consequence of electron-electron (e-e) interactions. 
Experimental signatures of TLL behavior include non-universal power laws in 
certain transport properties related to the tunneling density
of states, but many other observables also may reflect the non-Fermi 
liquid properties of a TLL. Experimental observations 
in semiconductor quantum wires\cite{yacoby} 
were explained by TLL parameters of the 
order of $g_c\approx 0.4$ to $0.5$, while the corresponding
parameter in SWNTs\cite{bockrath} was reported as 
$g_{c+}\approx 0.16$ to $0.3$.
Both values are significantly smaller than the respective
noninteracting value, $g=1$.

Very recently, graphene monolayers\cite{geim04,review} have
become available as a new realization of a 2DEG, albeit with
properties strikingly different from their semiconducting
counterparts. The kinetic energy of graphene close to one of
the Dirac ($K,K'$) points is described by a two-component chiral
Dirac-Weyl Hamiltonian\cite{Note,semenoff}
\begin{equation}\label{dirac}
H=v_{\rm F}\bm{\tau}\cdot({\bf p}-{\bf A})
\end{equation}
of massless relativistic particles moving at graphene's Fermi
velocity $v_{\rm F}\approx 10^6$~m/sec, instead of the usual
Schr\"odinger Hamiltonian ${\bf p}^{2}/2m^*$ (with effective
mass $m^*$). In Eq.~(\ref{dirac}), $\bm{\tau}$ denotes the
vector of Pauli matrices in sublattice (``pseudo-spin'') space,
while the physical spin as well as the valley ($K-K'$) degrees
of freedom are left implicit. Furthermore, we have allowed for
a static inhomogeneous orbital magnetic field perpendicular to
the graphene plane (the field components in the plane do not
affect orbital motion), ${\bf B}=B(x,y)\hat e_z$, which is
incorporated by minimal (Peierls) coupling in terms of the
corresponding vector potential ${\bf A}(x,y)$. This gives rise
to interesting magnetic barrier and magnetic confinement
effects.\cite{peeters93,review2,demartino,peetersnew}

It is well known, both theoretically\cite{muller} and
experimentally,\cite{singlesnake} that a magnetic-field gradient
can give rise to unidirectional 1D snake states. Such orbits
were recently studied theoretically in
graphene\cite{lambert,ghosh,cserti08} and in
SWNTs.\cite{snaketube} Snake states carry current along the
lines where the magnetic field changes sign, and hence is zero.
On a classical level, they can be understood as half-orbits of
different circulation sense (for $B>0$ and $B<0$), patched
together to form a unidirectionally propagating
orbit.\cite{footnew} Pairs of snake states running antiparallel
to each other are referred to as double snake
states.\cite{doublesnake} In many regards, double snake states
correspond to the standard left- and right-movers in 1D quantum
wires. For example, they should exhibit quantized conductance
in multiples of $4e^2/h$ (including spin and valley
degeneracies). Since the snake states are spatially separated,
this quantization should be robust: shallow impurities are not
expected to cause scattering between snake modes of opposite
directionality.

In this work, we address consequences of the long-ranged but
ultimately screened e-e
interactions within and between the counterpropagating snake
orbits in graphene magnetic wave\-guides. For a wide class of
experimentally relevant field profiles, we show that a TLL state
with broken Galilei invariance and extremely weak disorder
sensitivity can be realized. On a general level, the importance
of e-e interactions for the correct interpretation of
experimental data in graphene has recently been
stressed.\cite{bostwick} Theories describing e-e interaction
effects on the transport properties of electrons in graphene
have been proposed for strong homogeneous magnetic
fields\cite{FQHE} and for zero magnetic field.\cite{falko06} A
recent debate has discussed the question whether interacting
electrons in undoped graphene form a Fermi liquid or
not.\cite{vozmediano,khveshchenko06} Interactions are also
predicted to yield a TLL state in special graphene nanoribbons
with armchair edges.\cite{ribbons}

For the related case of interacting metallic SWNTs, the
effective low-energy theory predicts a four-channel TLL
state,\cite{swnt,swnt2} where the spin and valley degrees of
freedom give rise to the four channels. There is one charged
($c+$) channel, where the long-ranged e-e interactions play a
crucial role, while the three neutral channels are basically
insensitive to interactions. We will see that the situation in
a graphene magnetic waveguide is similar, and the parameter $g$
discussed below plays the role of the SWNT parameter $g_{c+}$.
Albeit the interaction does not spoil conductance quantization
in dc transport for adiabatically connected
reservoirs,\cite{safi} it nevertheless destroys the Fermi-liquid
character of the system. In fact, it leads to non-universal
power laws in the tunneling density of states, and to peculiar
ac transport and shot noise\cite{noise} properties at low
temperatures. These phenomena are appropriately described by
TLL theory.\cite{solyom,haldane,voit,gogolin,giamarchi} The
respective power-law exponents can be directly inferred from
Refs.~\onlinecite{swnt,swnt2} by simply replacing $g_{c+}$ with
our estimate for $g$, see Eq.~\eqref{gsemicl} below.

After introducing the model and the magnetic field profiles in
Sec.~\ref{sec2a}, the bandstructure is studied in Gaussian
approximation in what follows in Sec.~\ref{sec2b}. The
analytical bandstructure results are validated by comparing to
exact diagonalization results. The numerical diagonalization is
briefly discussed in the Appendix. The linearized bandstructure
for a double-snake state waveguide leading to TLL behavior is
then described in Sec.~\ref{sec3a}. The physics of a TLL is
governed by a dimensionless interaction parameter $g$, for which
general expressions in terms of certain velocities are derived
in Sec.~\ref{sec3b}. These velocities are obtained from
perturbative expressions for the ground-state energy, and yield
the analytical results for $g$ given in Sec.~\ref{sec4}. We
conclude in Sec.~\ref{sec5}. In Secs.~\ref{sec3} and
\ref{sec4}, to be specific, we focus on electron-like
excitations by assuming a positive value of the Fermi energy
$\varepsilon_{\rm F}$. Below, we often take units such that
$\hbar=v_{\rm F}=1$.

\section{Model and bandstructure}\label{sec2}

\subsection{Model}\label{sec2a}

In this paper, we consider magnetic fields $B=B(x)$,
guiding particles homogeneously along the $y$-direction.
This implies that the wavenumber $k$ along this direction is conserved.
Two-component eigenstates of the Dirac-Weyl Hamiltonian \eqref{dirac}
can then be written as
$\psi(x,y) \sim{\rm e}^{{\rm i}ky}({\phi_k(x), \chi_k(x)})^T$.
The vector potential can be chosen as ${\bf A}=A(x) \hat e_y$,
with the spinor components obeying
\begin{eqnarray}\label{spinors}
\displaystyle\left(\begin{matrix}0&-{\rm i}\partial_x-{\rm i}k+{\rm i}A(x)\cr
-{\rm i}\partial_x+{\rm i}k-{\rm i}A(x)&0\end{matrix}\right)
\left(\begin{matrix}\phi_k(x)\cr\chi_k(x)\end{matrix}\right)
\nonumber\\
\displaystyle=\varepsilon_k\left(\begin{matrix}\phi_k(x)\cr\chi_k(x)
\end{matrix}\right)\;. \qquad \qquad
\end{eqnarray}
Complex phases may be
chosen such that $\phi_k$ is real and $\chi_k$ purely imaginary.
After squaring, Eq.~(\ref{spinors}) can be cast into a
Schr\"odinger-like form for the upper Dirac component $\phi_k$,
\begin{equation}\label{schrodinger}
\left[-\partial_x^2+[k-A(x)]^2-B(x)-\varepsilon_k^2 \right]\phi_k(x)=0 \;.
\end{equation}
A similar equation holds for the lower component $\chi_k(x)$,
with the sign of the ``pseudo-Zeeman'' term $\sim B$ reversed.
Unless $\varepsilon_k=0$, Eq.~(\ref{spinors}) implies
\begin{equation}\label{eachhalf}
\int dx\;|\phi_k(x)|^2=\int dx\;|\chi_k(x)|^2=1/2.
\end{equation}
Note that in path-integral approaches to relativistic quantum
mechanics, in order to guarantee convergence of the Wiener
measure,\cite{hsquared} often the square of the Dirac
Hamiltonian \eqref{dirac} is considered. Path-integral
representations, on the other hand, allow for systematic
approximations, and therefore Eq.~(\ref{schrodinger}) is a
useful starting point for the Gaussian approximation, cf.\
Sec.~\ref{sec2b}, where the boundedness of the differential
operator appearing in Eq.~\eqref{schrodinger} is exploited for
either sign of $B(x)$, in the spirit of a saddle-point
approximation. Our method differs from WKB-type approaches
recently put forward to describe the electronic properties of
graphene.\cite{cserti08,ullmo} We note in passing that
massive Schr\"odinger particles obey a related equation as
Eq.~(\ref{schrodinger}), with quadratic momenta multiplied by
$1/(2m^*)$, in the same magnetic field profile; only the
pseudo-Zeeman term must be removed and, of course, the energy
$\varepsilon_k^2$ must be replaced by $\varepsilon_k$, i.e.\
hole and zero-energy states both disappear. As a consequence,
most of our conclusions also apply (at least qualitatively) to
magnetic waveguides based on traditional Schr\"odinger fermions.

We consider the class of magnetic field profiles given by
\begin{equation}\label{bix}
B(x)=\nu\omega_{\rm B}(\sqrt{\omega_{\rm B}}x)^{\nu-1}-B_0 \;.
\end{equation}
In our gauge, we thus have
\begin{equation}\label{aix}
A(x)=\omega_{\rm B}^{\frac{\nu+1}{2}}x^\nu-B_0 x \;.
\end{equation}
The index $\nu$ can describe rather different situations, but we
are only interested in $\nu$ being a natural number. For
instance, for $\nu=1$, we recover the homogeneous magnetic field
case, giving rise to the standard relativistic Landau levels.
For $\nu=2$, the profile \eqref{bix} instead describes a setup
with one snake state propagating along the $y$-direction, while
$\nu=3$ (or, more generally, all odd $\nu>1$) can give rise to a
double snake state geometry, where the background magnetic field
$-B_0$ allows for lines with $B=0$, and $\omega_{\rm B}$ sets
the inhomogeneity scale. Equation (\ref{schrodinger}) manifests
the electron-hole symmetry $\varepsilon_k\leftrightarrow
-\varepsilon_k$ of Eq.~(\ref{spinors}), with a zero-energy
eigenstate ($\varepsilon_k=0$ for all $k$) appearing whenever
$\nu$ is odd, but not for even $\nu$.\cite{indexthyrefs}

Equation \eqref{bix} qualitatively describes many situations of
experimental relevance, where typically smooth magnetic field
profiles are present. Of course, far away from the waveguide
defined by the snake states, the actual profile is different in
practice, but this does not significantly affect the TLL
discussed below. In fact, we have also analyzed step-like field
profiles, such as the ones described in Ref.~\onlinecite{ghosh},
with very similar results and conclusions. For $\nu=3$,
counterpropagating snake states are centered around $x= \pm d$
with $B(\pm d)=0$, leading to
\begin{equation}\label{ddef}
d = \frac{\sqrt{B_0}}{ \sqrt{3} \, \omega_{\rm B} } \;,
\end{equation}
such that $2d$ is the parallel distance between counterpropagating
snake states. In this configuration, a TLL can be realized,
and most of our analysis will deal with this case.

Below, we will ignore the Zeeman splitting due to the
interaction of the true electronic spin with the magnetic field
creating the waveguide. A simple estimate for a homogeneous
magnetic field already shows that this approximation is
justified in graphene. The physical Zeeman splitting
$\Delta_{\rm Z}=g_{\rm e}\mu_{\rm B}B$ amounts to $0.116$~meV
for $B=1$~Tesla, taking $g_{\rm e}=2$ and the free electron mass
$m_{\rm e}$ going into Bohr's magneton $\mu_{\rm B}$. This
value can be compared to the orbital splitting $\Delta_{\rm
orb}$ between subsequent levels --- in the language of
Eq.~\eqref{schrodinger}, this corresponds to a ``pseudo-Zeeman
splitting'' ---, with the result
\begin{equation}\label{zeemanpseudo} 
\frac{\Delta_{\rm Z}}{\Delta_{\rm orb}}\simeq 
\frac{\varepsilon_{\rm F}}{m_{\rm e}v_{\rm F}^2} \;,
\end{equation}
predicting that even for $\varepsilon_{\rm F}=1$~eV,
the Zeeman splitting is 50 times smaller than the orbital splitting.
In view of the smallness of the ratio \eqref{zeemanpseudo}, we will
neglect Zeeman terms in what follows. In any case, their effects
on the low-energy theory of interacting electrons in graphene
waveguides are standard, and could be included along the lines of
Refs.~\onlinecite{haldane,voit,gogolin,giamarchi}.

\subsection{Gaussian approximation} \label{sec2b}

Next we describe our analytical approach to the bandstructure
and the eigenfunctions. They allow for closed form
expressions of the TLL parameter $g$ in Sec.~\ref{sec4}. At
large $|k|$, when anharmonic contributions of the effective
potential appearing in Eq.~\eqref{schrodinger} are suppressed,
the Gaussian approximation becomes exact. To confirm the
accuracy of the analytical results, we have carried out
numerical diagonalizations of the matrix representing the
Schr\"odinger-like Hamiltonian (\ref{schrodinger}) in a complete
basis set. This is briefly described in the Appendix.

It is instructive to first study one of the two
counterpropagating snake modes individually. We therefore set
$B_0=0$ and $\nu=2$ in Eq.~\eqref{bix}, i.e.\
$B(\xi)=2\omega_{\rm B}\xi$ with $\omega_{\rm B}>0$, where we
introduce dimensionless lengths, $\xi=\sqrt{\omega_{\rm B}}\ x$,
and momenta, $\kappa=k/\sqrt{\omega_{\rm B}}$. The
Schr\"odinger version of this model has been studied
previously.\cite{muller} The single snake state is now centered
near $x=0$ with (positively or negatively charged) particles running in
the negative $y$-direction. The spectrum in this case is not
symmetric, $\varepsilon_k\ne \varepsilon_{-k}$. We then need to
discuss the effective potential appearing in
Eq.~(\ref{schrodinger}),
\[
V_{\nu=2}(\kappa,\xi)=(\xi^2-\kappa)^2-2\tau_z\xi\;,
\]
which depends on the sublattice component $\tau_z=\pm 1$ of the
wavefunction. Obviously, for any $\kappa$, $V_{\nu=2}$ is
invariant under the combined operation $\tau_z\to -\tau_z$ and
$\xi\to -\xi$, so that $|\phi_k(\xi)|=|\chi_k(-\xi)|$ in
Eq.~(\ref{spinors}). For $\kappa\to -\infty$, the minima of
$V_{\nu=2}$ approach $\kappa^2$ at $\xi=0$, so that
$|\phi_k(\xi)|=|\chi_k(\xi)|\sim e^{-\sqrt{-\kappa/2}\;\xi^2}$
and $\varepsilon_{k\to -\infty}\to -k$. This indicates that the
velocity reaches, up to subleading corrections, the negative of
the Fermi velocity. This is precisely the snake state, with
both components of the Dirac spinor localized near $x=0$. In
the other limit, $\kappa\to +\infty$, the minima of $V_{\nu=2}$
approach $-2\sqrt{\kappa}$ at $\xi=\tau_z\sqrt{\kappa}$. In
that case, the two sublattice components
$|\phi_k(\xi)|=|\chi_k(-\xi)|\sim
e^{-\sqrt{\kappa}[\xi-\sqrt{\kappa}]^2}$ are spatially separated
from one another and from the snake mode, provided the distance
exceeds the widths of these distributions. This result
suggests an interesting application as a ``sublattice filter'',
where the magnetic field leads to a spatial separation of
particles located on different sublattices. 
However, for the other $K$ point, the sublattice states are
exchanged, and in order to see such an effect, one would need to
have a valley-polarized system (i.e.\ a single $K$ point). The
corresponding energy in Gaussian approximation is
$\varepsilon_{k\to +\infty}\to 0$. This result cannot be
recovered using WKB-type approaches \cite{cserti08,ullmo} which
are more suited to describe higher excited states.
Interestingly, the positions $\xi=\pm\sqrt{\kappa}$ of these
states remain protected against the pseudo-Zeeman field
``inclination'' from Eq.~(\ref{schrodinger}), contrary to naive
expectation and in contrast to the non-zero shift found in any
of the excited states. Finally, also excited energies can be
estimated in this way. For example, the first excited level
is expected at $\varepsilon_k=2\omega_{\rm B}^{3/8}k^{1/4}$.

\begin{figure}
\includegraphics[width=0.45\textwidth]{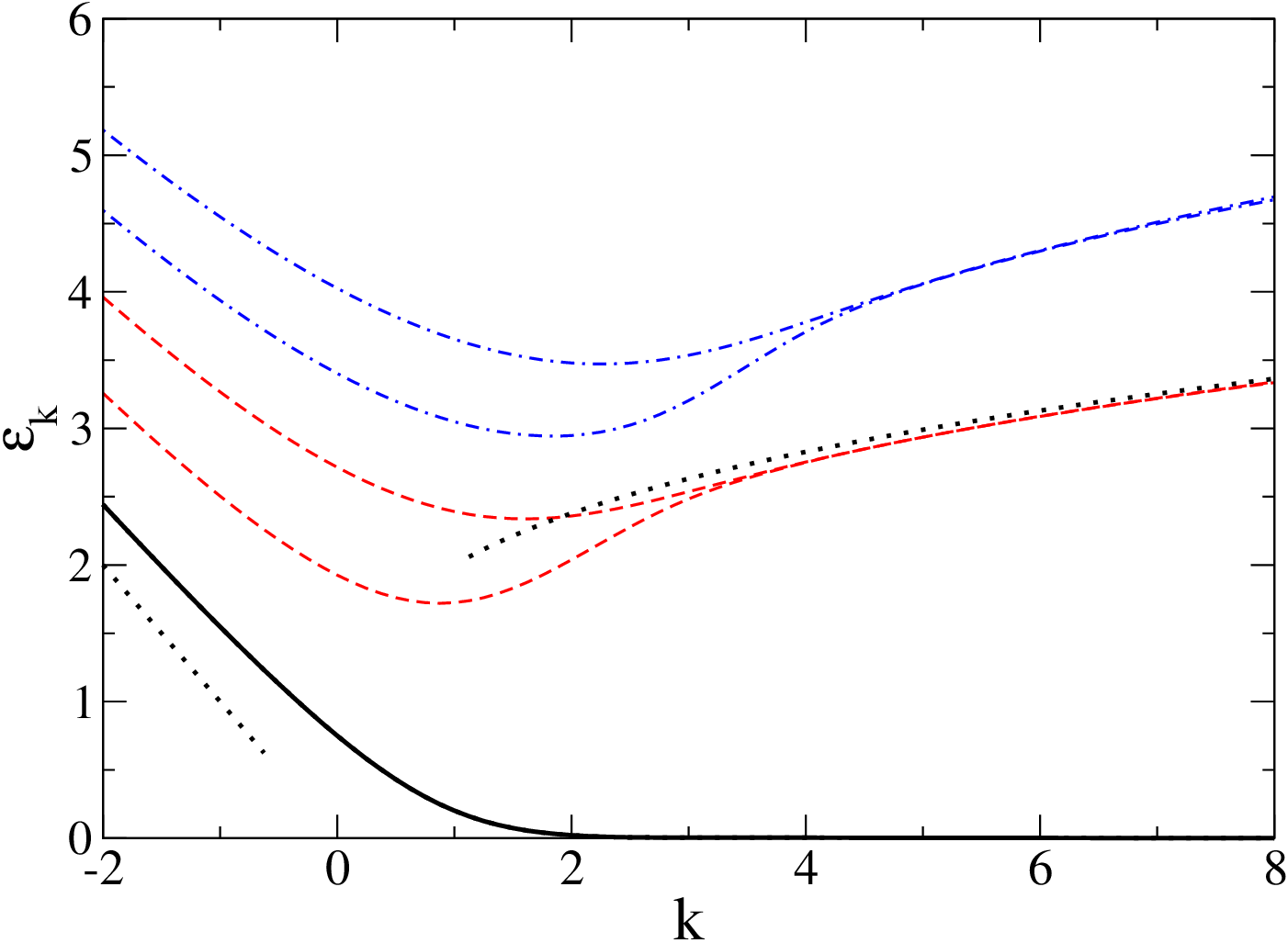}
\caption{\label{fig1} (Color online) Electron-like energy
eigenvalues $\varepsilon_k$ versus momentum $k$ (both in units
of $\sqrt{\omega_B}$), as obtained by numerical diagonalization
of Eq.~\eqref{schrodinger} for a magnetic field profile with
$\nu=2$ and $B_0=0$ in Eq.~\eqref{bix}. The solid (black) curve
denotes the lowest positive-energy eigenstate, and the dashed
(red) and dash-dotted (blue) curves give the next two pairs of
excited states. The dotted curve indicates the limiting
snake-state dispersion $\varepsilon_k=-k$ for $k<0$, and the 
result in Gaussian approximation for the first excited energy
band, $\varepsilon_k=2\omega_{\rm B}^{3/8}k^{1/4}$, for $k>0$.
Note that for the lowest state, $\varepsilon_{k\to +\infty}$
approaches zero energy,\cite{lambert,ghosh} in agreement
with the Gaussian approximation.}
\end{figure}

\vspace{1cm}

\begin{figure}
\includegraphics[width=0.44\textwidth]{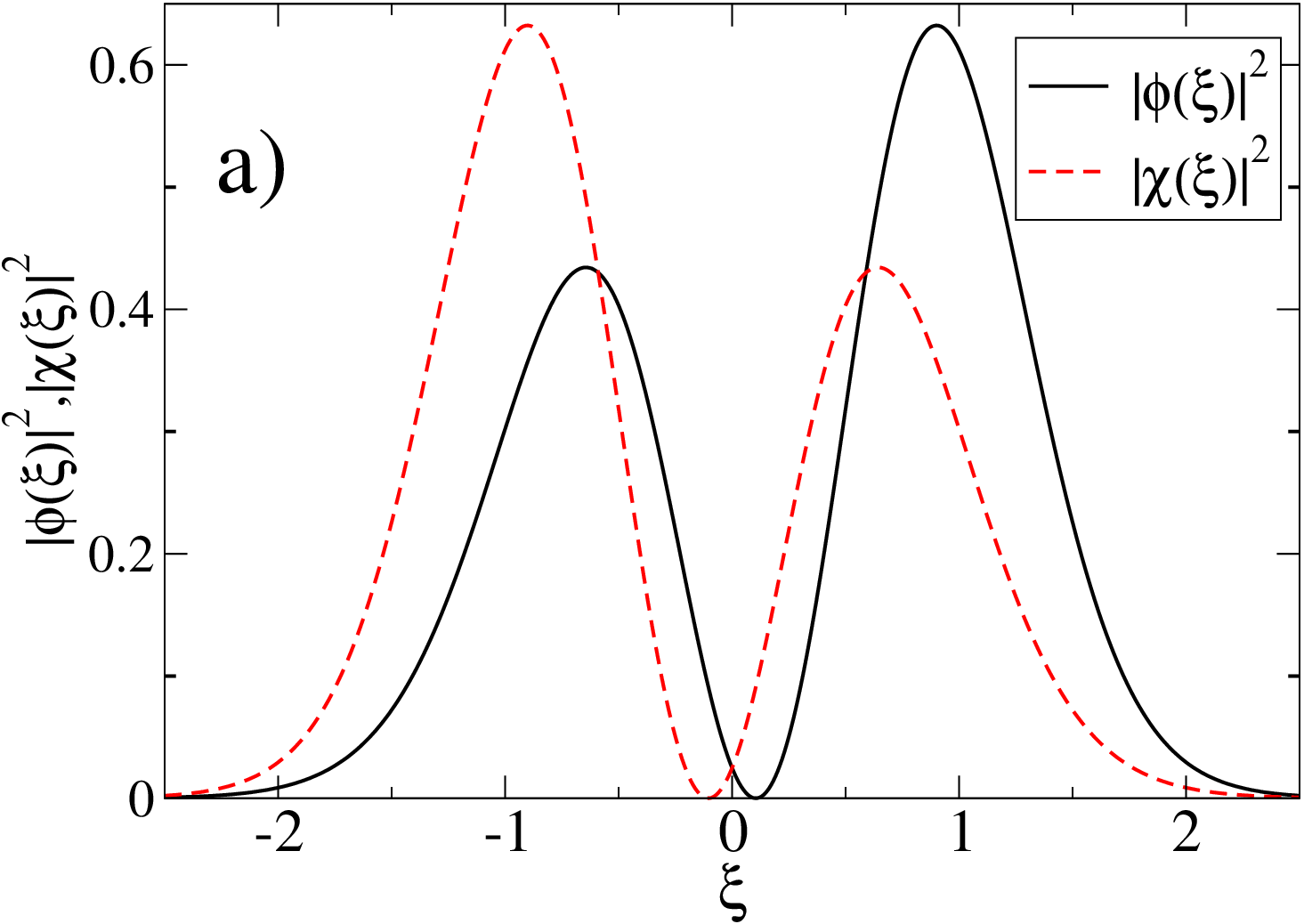}

\vspace{1cm}

\includegraphics[width=0.44\textwidth]{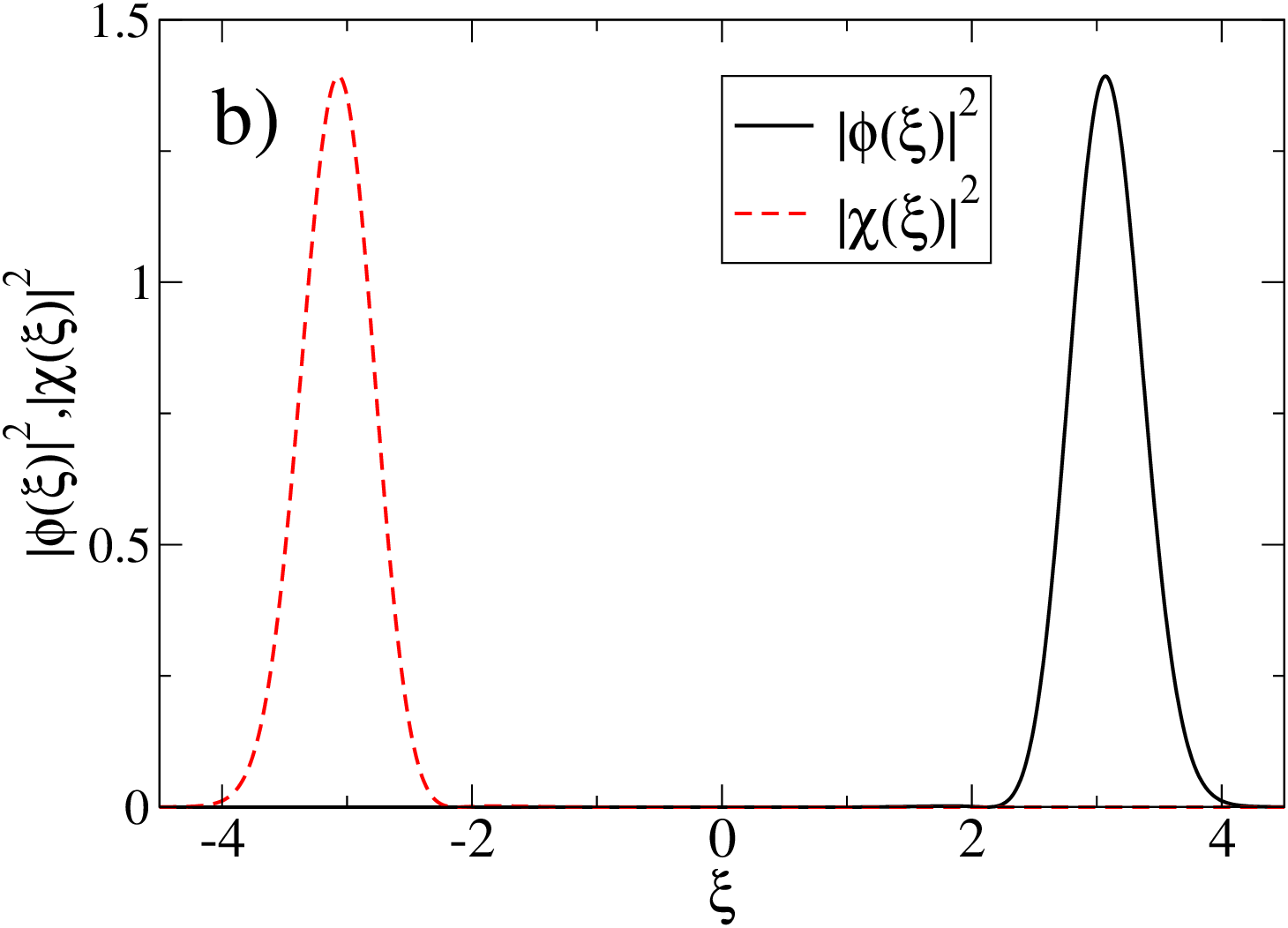}
\caption{\label{fig2}
(Color online) Numerical diagonalization results for the
probability densities of the spinor components
$|\phi_{\kappa}(\xi)|^2$ (black solid curve) and
$|\chi_{\kappa}(\xi)|^2$ (red dashed curve) of the lowest
eigenstate to positive energy of Eq.~(\ref{spinors}) with
$\nu=2$ and $B_0=0$. (a) is for momentum $\kappa=-2$, and (b)
for $\kappa=8$.}
\end{figure}

\vspace{1cm}

In effect, we then arrive at a picture where snake (near $x=0$)
and ``bulk'' modes (near $x=\pm \omega_{\rm B}^{-3/4}\sqrt{k}$)
will develop. We here distinguish ``snake'' and ``bulk'' modes
by their respective group velocities. Fig.~\ref{fig1} clearly
demonstrates how the eigenstates evolve from snake states (at
$k\to-\infty$ with $\partial_k\varepsilon_k=-1$) into bulk modes
at $k\to+\infty$. In Fig.~\ref{fig2} the corresponding
metamorphosis is displayed of a (single) snake state at
sufficiently negative $k$ centered around $\xi=0$, see
Fig.~\ref{fig2}(a), into a bulk state, residing increasingly far
away from $\xi=0$ with increasing $k>0$. For $\kappa=8$,
cf.\ Fig.~\ref{fig2}(b), the state is located near
$\xi=2.83\tau_z$ as expected.

\begin{figure}
\includegraphics[width=0.44\textwidth]{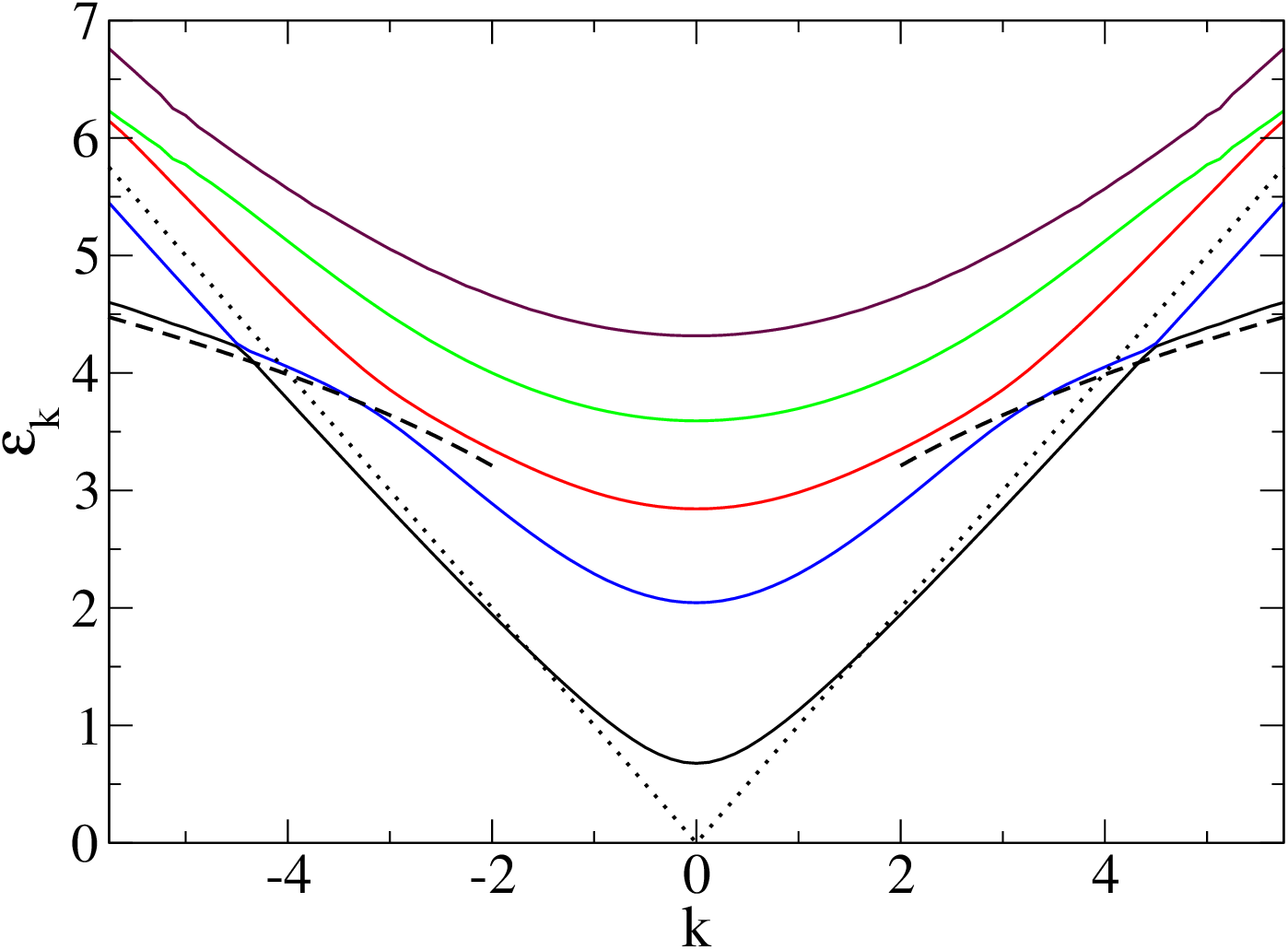}
\caption{\label{fig3} (Color online) Same as Fig.~\ref{fig1} but
for $\nu=3$ and $B_0= 1.526\omega_{\rm B}$. Full curves (in
different colors) give the numerical diagonalization results for
the eigenenergies. When the Fermi level intersects only the
lowest (black) curve, one has precisely one (spin- and
valley-degenerate) left- and right-moving state in the
waveguide. This leads to a TLL state. The dotted curve denotes
$\varepsilon_k=|k|$, and the dashed curve gives the 
estimate \eqref{sce} for the lowest dispersing
energy band in the Gaussian approximation.}

\vspace{1cm}

\end{figure}

Let us now turn to the magnetic field profile with $\nu=3$ in
Eq.~(\ref{bix}), which should exhibit double snake states of
opposite directionality due to the existence of two zeros of
$B(x)$. To some extent, we now have two copies of the above
single-snake state situation at a distance $2d$, cf.\
Eq.~\eqref{ddef}. Since $B(x)=B(-x)$, the dispersion relation
is now symmetric, $\varepsilon_k=\varepsilon_{-k}$. At
$k\to\pm\infty$, we thus anticipate the coexistence of snake and
bulk modes. In addition, an exact zero-energy eigenstate,
$\varepsilon_k=0$, must now occur as a consequence of the index
theorem.\cite{indexthyrefs} The potential entering
Eq.~(\ref{schrodinger}) is
\begin{equation}\label{vnu3}
V_{\nu=3}(\kappa,\xi)=(\xi^3-b_0\xi-\kappa)^2-\tau_z(3\xi^2-b_0) \;,
\end{equation}
where $b_0=B_0/\omega_{\rm B}$. This potential is invariant
under a simultaneous sign change of $\kappa$ and $\xi$,
transforming left- into right-movers and vice versa. Similar as
for $\nu=2$, we can obtain energies in Gaussian
approximation. The lowest energy, only for $\tau_z=+1$, equals
zero at large $|k|$. This describes the zero-energy state,
present for any $k$ at odd $\nu$, cf.\ Sec.~\ref{sec2a} and the
Appendix. The first excited (positive) energy is approximated as
\begin{equation}\label{sce}
\varepsilon_k=\sqrt{6}|\omega_{\rm B}k|^{1/3}+
B_0/[\sqrt{6}|\omega_{\rm B}k|^{1/3}]\;.
\end{equation}
This result is included to the numerically obtained spectra, see
Fig.~\ref{fig3}. The corresponding eigenstates are localized
around
$\xi=(|\kappa|^{1/3}+b_0/3|\kappa|^{1/3}+\tau_z/3|\kappa|){\rm
sgn}(\kappa)$, i.e.\ increasingly deep in the system's bulk with
increasing $|k|$. Fig.~\ref{fig3} shows a typical spectrum
obtained from numerical diagonalization. Depending on the
slopes $\partial_k\varepsilon_k$ at large $|k|$, one type of
bands goes like $\varepsilon_k\simeq \pm v_{\rm F} |k|$,
corresponding to the counterpropagating snake states centered
around $x=\pm d$, see Eq.~\eqref{ddef}. Snake states move with
the Fermi velocity of graphene at large $|k|$, irrespective of
the magnetic field profile.\cite{ghosh,lambert,park08} The other
bands in Fig.~\ref{fig3} exhibit smaller slopes at large $|k|$.
The corresponding states are localized increasingly further away
from the center of the wire at increasing $|k|$, and we thus
call them again ``bulk'' modes.  Due to their different
slopes, bands of different types should cross, and
Fig.~\ref{fig3} indeed reveals avoided intersections. Such
avoided level crossings can be attributed to some residual
hybridization between snake and bulk modes. They become
successively less important as the bulk state's center moves
away from the snake state with increasing $|k|$.

\begin{figure}
\includegraphics[width=0.45\textwidth]{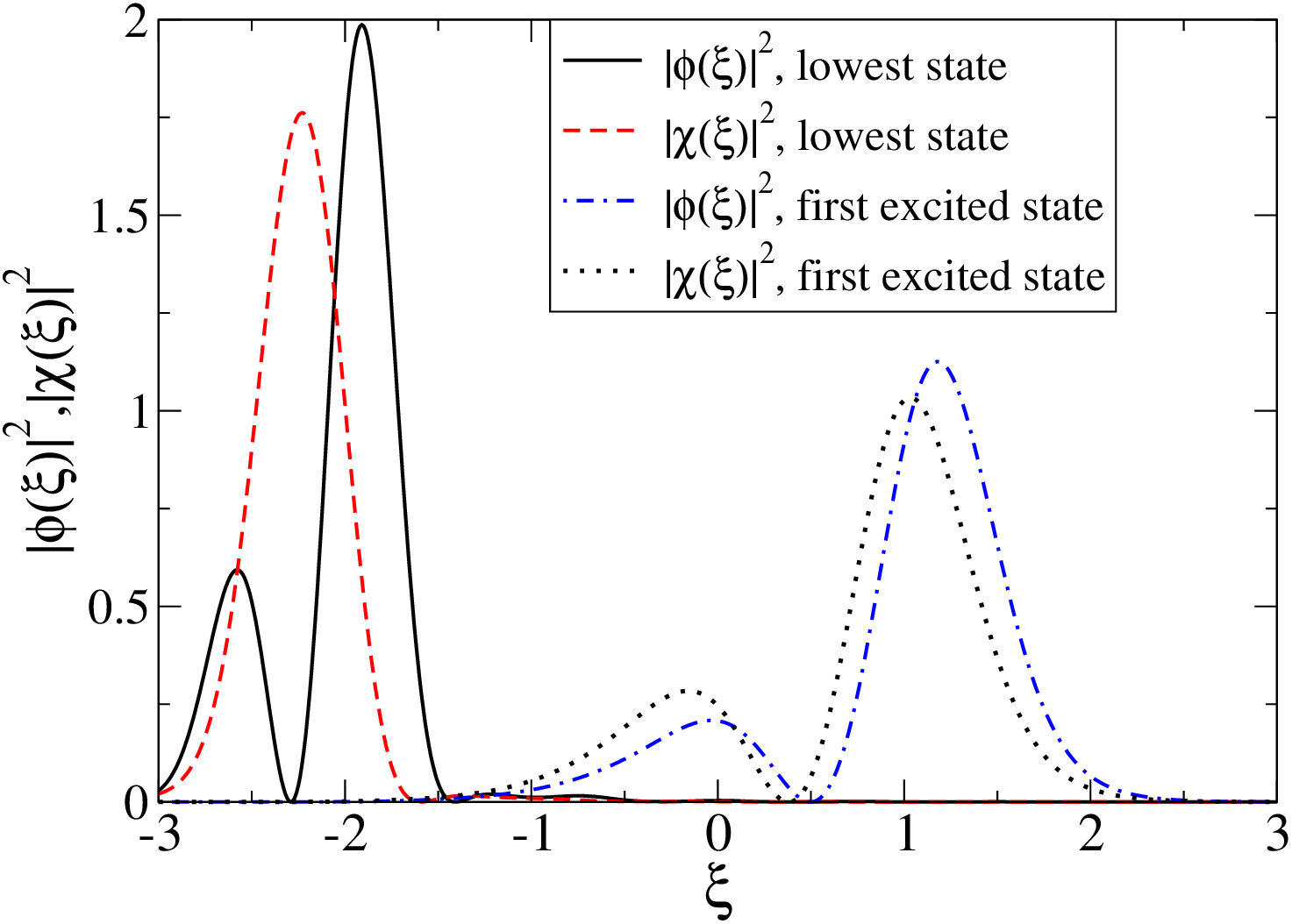}
\caption{\label{fig4} (Color online) Same as Fig.~\ref{fig2} but for
a magnetic field \eqref{bix} with $\nu=3$ and $B_0=1.525\omega_B$. Shown
are the two lowest positive-energy eigenstates with $\kappa=-6$.}
\end{figure}

Fig.~\ref{fig4} validates the behavior of the two types of
eigenstates for the double-snake $\nu=3$ situation with
$\kappa=-6$. In view of Fig.~\ref{fig3}, this value of
$|\kappa|$ is just beyond one of the (narrower) avoided
crossings, so that the lowest positive-energy state, exhibiting
a moderate slope, must be identified as a bulk state. This is
indeed confirmed in Fig.~\ref{fig4}, where the densities of both
components of this state are seen. Gaussian approximation
predicts its center to be at $\xi=-2.1+0.06\tau_z$, which is
nicely confirmed by comparing to exact diagonalization results.
The first excited state at $\kappa=-6$ has $\partial_k
\varepsilon_k\simeq -1$, and therefore is classified as snake
state. Indeed, as seen in Fig.~\ref{fig4}, both pseudo-spin
components of this state reside near $\xi=0.7$, where
$B(\xi)=0$. Note that, according to Eq.~(\ref{nk}), densities
for odd $\nu$ are mirror-symmetric under simultaneous sign
reversal of $\kappa$ and $\xi$, so that the $\kappa=+6$ state is
centered at $\xi=-0.7$. Finally, we note that all densities in
Figs.~\ref{fig2} and \ref{fig4} are well approximated by
superpositions of suitable Gaussians, in accordance with our 
Gaussian description.

\section{Interaction effects} \label{sec3}

\subsection{Waveguide model} \label{sec3a}

When considering a 1D graphene magnetic waveguide with
counterpropagating snake states at low energy scales, see
Sec.~\ref{sec2b}, the question arises whether {\sl interacting}\
Dirac fermions in such a waveguide belong to the TLL
universality class.\cite{solyom,haldane,voit,gogolin,giamarchi}
As always happens in 1D, Landau quasiparticles will be destroyed
by any non-zero e-e interaction strength,\cite{larkin74} but
whether the resulting non-Fermi liquid is a TLL remains to be
shown. Indeed, this expectation is corroborated by the
analogous situation in a quantum Hall bar,\cite{qheref,chang}
where, as a consequence of the long-ranged Coulomb interaction
between different edge states, a TLL with spatially separated
left- and right-moving edge states emerges.\cite{UZ}

We start with a discussion of the relevant single-particle
bandstructure. For a magnetic waveguide with $\nu=3$ and
$B_0>0$ in Eq.~\eqref{bix}, the lowest positive-energy subband
$\varepsilon^{(1)}_k=\varepsilon^{(1)}_{-k}$ has left- and
right-going snake states near $x=\pm d$, see Eq.~\eqref{ddef},
that essentially move at velocity $\pm v_{\rm F}$. We assume
that the Fermi level intersects these states at $\pm k_{\rm F}$,
and that all other states are energetically sufficiently far
away. This bandstructure was discussed in detail in
Sec.~\ref{sec2b}, see Fig.~\ref{fig3}. Using second
quantization, the kinetic energy is then described by
\begin{equation}\label{free}
H_0=\sum_{k}\varepsilon^{(1)}_k c_{k}^\dagger c_{k}^{}\;,
\end{equation}
where $c^\dagger_k$ ($c_k^{}$) creates (annihilates) a Dirac
quasiparticle with momentum $k$, and spin and valley indices are
kept implicit. The electron field operator for waveguide length
$L$ along the $y$-direction is thus written as
\begin{equation}
\Psi(x,y) = \frac{1}{\sqrt{L}} \sum_k e^{{\rm i}ky}
\left(\begin{array}{c}  \phi_k(x) \\ \chi_k(x)\end{array} \right) \ c_k\;.
\end{equation}
The crucial parameters characterizing the TLL are certain
velocities.\cite{voit} In the noninteracting case, Dirac
particles move at velocity\cite{ghosh}
\[
v_{\rm F}\langle\tau_y\rangle_k=2v_{\rm F}{\rm Im}\int{\rm
d}x\ \phi_k^*(x)\chi_k^{}(x)=\partial_k\varepsilon_k \equiv v_k \;,
\]
given by the slope of the energy dispersion, just as for
Schr\"odinger particles. Assuming that $\varepsilon_{\rm F}$ is
sufficiently far away from both the band bottom of
$\varepsilon^{(1)}_k$ and from the next-higher energy band, we
linearize $\varepsilon^{(1)}_k$ about the two Fermi points
$k=\pm k_{\rm F}$, yielding velocities $\pm v_{k_{\rm F}}$.
This also separates right ($k>0$) from left ($k<0$) movers in
Eq.~(\ref{free}), and implicitly defines the standard bandwidth
cutoff around the Fermi level used in TLL theory.

\subsection{Tomonaga-Luttinger liquid}\label{sec3b}

Next we incorporate e-e interactions within an effective
low-energy theory. We consider the pair interaction potential\cite{whlkahm}
\begin{equation}\label{coulomb}
W({\bf x}_1,{\bf x}_2)=\frac{e^2}{\kappa_0} 
\left( \frac{1}{|{\bf x}_1-{\bf x}_2|}
-\frac{1}{\sqrt{({\bf x}_1-{\bf x}_2)^2+4D^2}} \right)
\end{equation}
between electrons at coordinates ${\bf x}_1=(x_1,y_1)$ and ${\bf
x_2}=(x_2,y_2)$. This form specifically accounts for screening
by metal gates positioned at some distance $D$ from the
waveguide. Its strength is governed by the dimensionless ``fine
structure constant'' $\alpha=e^2/[\kappa_0 \hbar v_{\rm F}]$,
which basically depends only on the dielectric constant
$\kappa_0$. For typical substrate materials, one has values
$\kappa_0\approx 1.4$ to $4.7$, resulting in $\alpha\approx 0.6$
to 2.\cite{swnt2,alicea,noWC} For graphene, both the kinetic as
well as the Coulomb energy scale ($\sim\sqrt{n}$ for particle
density $n$) in the same way.\cite{noWC} The resulting weak
tunability of the e-e interaction strength in graphene is in
stark contrast to the situation in semiconductors, where $n$
allows to alter the relative strength of Coulomb interactions
over orders of magnitude.

When constructing a low-energy theory for interacting Dirac
fermions in the double snake state waveguide of
Sec.~\ref{sec3a}, the resulting 1D e-e interaction processes can
be classified as forward-scattering and backscattering
processes.\cite{solyom,haldane,voit,gogolin,giamarchi} The
spatial separation of the unidirectional snake states here
implies a strong suppression of e-e backscattering processes,
where the relevant couplings are exponentially small in the
parameter $k_{\rm F}d \gg 1$. In the following, we discuss the
regime
\begin{equation} \label{regime}
k_{\rm F} D \gg k_{\rm F} d \gg 1\;,
\end{equation}
where backscattering processes are negligible. This situation
is reminiscent of the SWNT case,\cite{swnt2} where one, however,
finds only an algebraic suppression of the backscattering
couplings with increasing SWNT radius. We then only need to
include forward-scattering processes, see also
Ref.~\onlinecite{swnt2}, and arrive at a four-channel TLL model.
The three neutral sectors involve spin and valley degrees of
freedom, and are decoupled from each other and from the charge
sector (spin-charge separation). For not too strong
interactions, as expected in graphene, the neutral sectors will
remain basically unaffected by interactions, with their velocity
parameters (almost) equal to $v_{k_{\rm F}}$, see also
Ref.~\onlinecite{epl}. This implies that the TLL parameters for
the three neutral sectors are just given by the noninteracting
value ($g=1$). In the following, we then focus on the charge
($c+$) sector only.

The resulting TLL is most conveniently described by Abelian
bosonization.\cite{haldane,voit,gogolin,giamarchi,swnt2}
For the $c+$ sector, the resulting Hamiltonian is
\begin{equation}\label{boseTLL}
H_{c+}= \frac12 \int dy\;\left [ v_{\rm J} [\partial_y\Theta(y)]^2+v_{\rm
N}[\partial_y\Phi(y)]^2\right]\;,
\end{equation}
with bosonic fields subject to the algebra
$[\Phi(y),\partial_y\Theta(y)]={\rm i}\delta(y-y')$. Equation
\eqref{boseTLL} reflects the fact that density waves (in
contrast to quasiparticles) are undamped in a
TLL.\cite{larkin74} With Eq.~(\ref{boseTLL}) and the usual
bosonized form of the Fermi operators,\cite{gogolin} almost any
observable of physical interest can be determined exactly at low
energies and long wavelengths, where the TLL model applies. In
general, $v_{\rm J}<v_{\rm N}$ can deviate from $v_{k_{\rm F}}$
as a result of the repulsive e-e interactions. Interaction
physics is thus encoded in $v_{\rm J}$ and $v_{\rm N}$. These
velocities determine the dimensionless TLL interaction parameter
$g\equiv g_{c+}$ and the plasmon velocity $v$ according to
\cite{haldane}
\begin{equation}\label{exponent}
g=\sqrt{v_{\rm J}/v_{\rm N}}\;, \quad
v=\sqrt{v_{\rm J}v_{\rm N}}\;.
\end{equation}
In principle, both parameters are experimentally accessible, e.g.\
through the tunneling density of states,\cite{bockrath} momentum-resolved
tunneling,\cite{yacoby} or via plasmon propagation times.\cite{ernst97}

The velocities $v_{\rm J}$ and $v_{\rm N}$ can be extracted
in an elegant and exact manner from thermodynamic relations,\cite{voit,wh}
\begin{equation}\label{thermodyn}
v_{\rm N}=\frac{\pi}{4L}\frac{\partial^2E_0}{\partial k_0^2} \;, \quad
v_{\rm J}=\frac{\pi}{4L}\frac{\partial^2E_0}{\partial\delta^2}\;,
\end{equation}
provided the fully interacting ground-state energy density
$E_0/L$ for fixed left $\left (-k_{\rm F}^{(-)}\right)$ and
right $\left(k_{\rm F}^{(+)}\right)$ Fermi momenta is known,
where $\delta=\left(k^{(+)}_{\rm F}-k^{(-)}_{\rm F}\right)/2$
and $k_0= \left(k_{\rm F}^{(+)}+ k_{\rm F}^{(-)}\right)/2$. The
derivatives in Eq.~(\ref{thermodyn}) are evaluated at $\delta=0$
and $k_0=k_{\rm F}$, and the energy $E_0$ includes the spin and
valley degrees of freedom. Clearly, $v_{\rm N}$ is proportional
to the compressibility, and when Galilei invariance is realized,
$v_{\rm J}=v_{k_{\rm F}}$ is unchanged by interactions.
However,  as we shall see below, this symmetry is {\sl
not}\ obeyed here due to the periodic superstructure imposed by
the snake orbit.

\section{TLL parameter}\label{sec4}

Unfortunately, exact results for $E_0$ are known only for a
limited number of integrable models, such as the Hubbard model
\cite{liebwu} or the Sutherland model.\cite{kawakami91} Even
then, one still has to numerically solve coupled pairs of
integral equations to access $E_0$, and hence the velocities
$v_{\rm J,N}$ in Eq.~\eqref{thermodyn}. In actual calculations
of $v_{\rm J}$ and $v_{\rm N}$ for nonintegrable models (which
is the case here), one has to resort to
approximations.\cite{whlkahm,lkwh}

\subsection{Perturbation theory}\label{sec4a}

We now use perturbation theory to obtain $E_0$, and thus the
velocities \eqref{thermodyn}, for relatively weak interactions.
This approximation is almost exclusively used in the literature
in order to obtain estimates for the TLL parameters of generic
interacting 1D fermion systems. In that case, the ground-state
energy can be split into three terms,
\begin{equation}
E_0=E_{\rm kin}+E_{\rm Hartree}-E_{\rm Fock}\;,
\end{equation}
from which several contributions to the
susceptibilities (\ref{thermodyn}) follow.
Accounting for spin and valley degeneracy, we have
\begin{equation}\label{ekinkk}
\frac{\partial^2 E_{\rm kin}}{\partial k_{0}^2}=
\frac{\partial^2E_{\rm kin}}{\partial\delta^2}=
\frac{2L}{\pi}(\partial_k\varepsilon^{(1)}_{k_{\rm F}}-\partial_k
\varepsilon^{(1)}_{-k_{\rm F}})=
\frac{4L}{\pi} v_{k_{\rm F}}\;,
\end{equation}
as expected for noninteracting quasiparticles
with velocities $\pm v_{k_{\rm F}}$.

{}From the Hartree interaction term, we then obtain
\begin{equation}\label{eharkk}
\frac{\partial^2 E_{\rm Hartree}}{\partial k_{0}^2}=
\frac{L}{\pi^2}\Bigl[\eta(k_{\rm F},k_{\rm F})+
\eta(k_{\rm F},-k_{\rm F})+{\cal R}\Bigr]\;,
\end{equation}
where, including again both spin and valley components, we define
\begin{eqnarray}
\eta(k,k')&=&\eta(k',k)= 8\int dx \int dx'\;n_k(x) n_{k'}(x')
\nonumber \\  &\times& \alpha\ln\sqrt{1+[2D/(x-x')]^2}\;, \label{etadef}
\end{eqnarray}
with the particle density for wavevector $k$ at location $x$,
\begin{equation}\label{nk}
n_k(x)=|\phi_k(x)|^2+|\chi_k(x)|^2=n_{-k}(-x) \;.
\end{equation}
Note that $\int dx\;n_k(x)=1$, see Eq.~\eqref{eachhalf}.
Furthermore, we have introduced the quantity
\begin{eqnarray}\label{rdef}
{\cal R}&=&\int_{-k_{\rm F}}^{k_{\rm F}} dk \;
\partial_{k_{\rm F}}[\eta(k,k_{\rm F})+\eta(k,-k_{\rm F})]\nonumber\\
&=&2\int_0^{k_{\rm F}} dk\;\partial_{k_{\rm F}}
[\eta(k,k_{\rm F})+\eta(k,-k_{\rm F})]\;.
\end{eqnarray}
The second equalities in Eqs.~(\ref{nk}) and (\ref{rdef}) are
valid for all odd $\nu>1$. In Eq.~\eqref{etadef}, we have used
that only long wavelengths $q\to 0$ are important in $E_{\rm
Hartree}$. To see this, consider the one-sided Fourier
transform of Eq.~\eqref{coulomb},
\begin{eqnarray}\nonumber
W(q,x)&=&\alpha\int dy\; e^{{\rm i}qy}\left(
\frac{1}{\sqrt{x^2+y^2}} -\frac{1}{\sqrt{x^2+y^2+4D^2}}\right) \\
\label{wqx}
&=&\alpha\left [K_0(|qx|)-K_0(|q|\sqrt{4D^2+x^2}) \right] ,
\end{eqnarray}
implying that $W(q\to 0,x)\simeq \alpha\ln\sqrt{1+[2D/x]^2}$.
With these definitions, we obtain in a similar manner
\begin{equation}\label{ehardd}
\frac{\partial^2 E_{\rm Hartree}}{\partial\delta^2}=
\frac{L}{\pi^2}\Bigl[\eta(k_{\rm F},k_{\rm F})-
\eta(k_{\rm F},-k_{\rm F})+{\cal R}\Bigr]\;.
\end{equation}

In view of Eqs.~(\ref{exponent}) to (\ref{ehardd}), only the
magnitude of $\eta(k_{\rm F},-k_{\rm F})$ yields a nontrivial
contribution to $g$ at this level of approximation. Within the
$g$-ology terminology,\cite{solyom} we may identify this term
with $g_2$, measuring the strength of forward scattering between
particles of different directionality (henceforth referred to as
chirality). On the other hand, scattering between
equal-chirality particles is described by $g_4$, identified here
as $\eta(k_{\rm F},k_{\rm F})$. However, according to
Eqs.~(\ref{eharkk}) and (\ref{ehardd}), an additional term
${\cal R}$ is present, which effectively modifies $g_4$. In 1D
quantum wires with continuous (Galileian) translational
invariance, both $n_k(x)$ and $\eta(k,k')$ are independent of
$k,k'$, and hence ${\cal R}=0$ and $\partial_{\delta}^2E_{\rm
Hartree}=0$. In that case, the TLL parameter $g$ depends only
on the zero-momentum Fourier component of the interaction,
$W(q\simeq 0) \approx \alpha\ln(D/d)$, where $d$ is of the order
of the wire width.

Similar (though slightly more involved) expressions can be found
for the Fock contributions to Eq.~(\ref{thermodyn}), which
are of the order $W(q\simeq 2k_{\rm F})$. Using Eq.~\eqref{regime},
this amplitude can be estimated as
\[ 
W(2k_{\rm F})\simeq \alpha\sqrt{\frac{\pi}{8k_{\rm F}d}}e^{-4k_{\rm F}d}\;,
\] 
which is parametrically smaller than the Hartree amplitude. 
Similar to backscattering contributions, Fock contributions
can thus safely be neglected against the Hartree terms for the
parameter regime \eqref{regime}.

\subsection{TLL parameter estimate} \label{sec4b}

In order to estimate the magnitude of the above terms, in
particular of $\eta(k_{\rm F},-k_{\rm F})$, it is necessary to
have some handle on the unperturbed wavefunctions $\phi_k(x)$
and $\chi_k(x)$, together with the resulting densities $n_k(x)$.
We approximate their density profiles as
\begin{equation}\label{nkappa}
n_{\kappa}(\xi)\simeq\frac{(12b_0\kappa^2)^{1/8}}{\sqrt{\pi}}
e^{-(12b_0\kappa^2)^{1/4}[\xi+{\rm sgn}(\kappa)\sqrt{\omega_{\rm
B}}d]^2}\;.
\end{equation}
Note that the true densities describing snake orbits, see
Fig.~\ref{fig4}, are somewhat more complicated, with a
double-peak shape. However, the simplified single-Gaussian form
in Eq.~\eqref{nkappa} captures the essential physics and allows
for analytical progress.

Given Eq.~(\ref{nkappa}), introducing the lengthscale
\begin{equation}\label{ldef}
\lambda = \left(\frac{3}{4}B_0\omega_{\rm B}^2k_{\rm F}^2 \right)^{-1/8}\;,
\end{equation}
and using Eq.~(\ref{etadef}), we are now in a position to estimate
\begin{eqnarray}\nonumber
\eta(k_{\rm F},-k_{\rm F}) &\simeq & 8\alpha\ln(D/d)\;,\\
\label{g41}
\eta(k_{\rm F},k_{\rm F})&\simeq & 8\alpha\ln\left(e^{C/2}
D/\lambda\right)\;,
\end{eqnarray}
assuming $D\gg d\gg\lambda$, see Eq.~\eqref{regime}. Here $C=
0.577\ldots$ is the Euler constant. We see that the ratio
$\eta(k_{\rm F},-k_{\rm F})/\eta(k_{\rm F},k_{\rm F})$
approaches unity for $D\to\infty$ as in Galilei-invariant 1D
wires, where in addition one also has ${\cal R}=0$. In order to
estimate ${\cal R}$, we first observe that in Eq.~(\ref{rdef}),
the $\eta(k,-k_{\rm F})$ term is suppressed by a factor
$e^{-8(d/\lambda)^2}$ as compared to $\eta(k,k_{\rm F})$. This
factor becomes small in the regime \eqref{regime}, and we can
therefore approximate ${\cal R}\simeq 2\int_0^{k_{\rm F}}
dk\;\partial_{k_{\rm F}} \eta(k,k_{\rm F}),$ which can be
calculated in closed form for the density profile
(\ref{nkappa}),
\begin{eqnarray} \nonumber
{\cal R}&\simeq &8\alpha\left[\sqrt{2}(c_1+2c_2)+4(\sqrt{2}c_1-1)
\ln(D/\lambda)\right]\\
&\approx&\alpha[5.73-2.60\ln \left(D/\lambda\right)]\;.\label{calr}
\end{eqnarray}
Employing the incomplete Euler Beta function, we find
\begin{equation}
c_1= \frac{1}{3}\left(4\sqrt{2}-\int_0^1dt\;
\frac{\sqrt{1+t}}{t^{3/4}}\right) \simeq 0.6496 
\end{equation}
and
\begin{equation}
c_2=\int_0^1 dt \ \frac{t^{1/4}}{\sqrt{1+t}}
\ln\left[ \frac{1+t}{1+\sqrt{t}}\right]\simeq -0.07171.
\end{equation}
Remarkably, ${\cal R}$ decreases with increasing $D/\lambda$,
changes sign at $D\simeq 9.02\lambda$, and then continues to
decrease logarithmically. Although asymptotically smaller by a
factor $4(\sqrt{2}c_1-1)\simeq 0.326$ than the leading
contribution (\ref{g41}) to $g_4$, the ${\cal R}$-term is
important for quantitative estimates of the TLL parameter. 
For example, when ${\cal R}<0$, standard expressions (without
${\cal R}$) would overestimate $g$, pretending too weak
interaction effects. The usual expressions are based on Galilei
invariance, which is broken in the present system due to the
periodic superstructure imposed on the 1D wire by the
snake orbits. It is not obvious to us how
standard estimates to $g_4$, starting from the microscopic
interaction \eqref{coulomb}, would recover the ${\cal
R}$-contribution.

\begin{figure}
\includegraphics[width=0.4\textwidth]{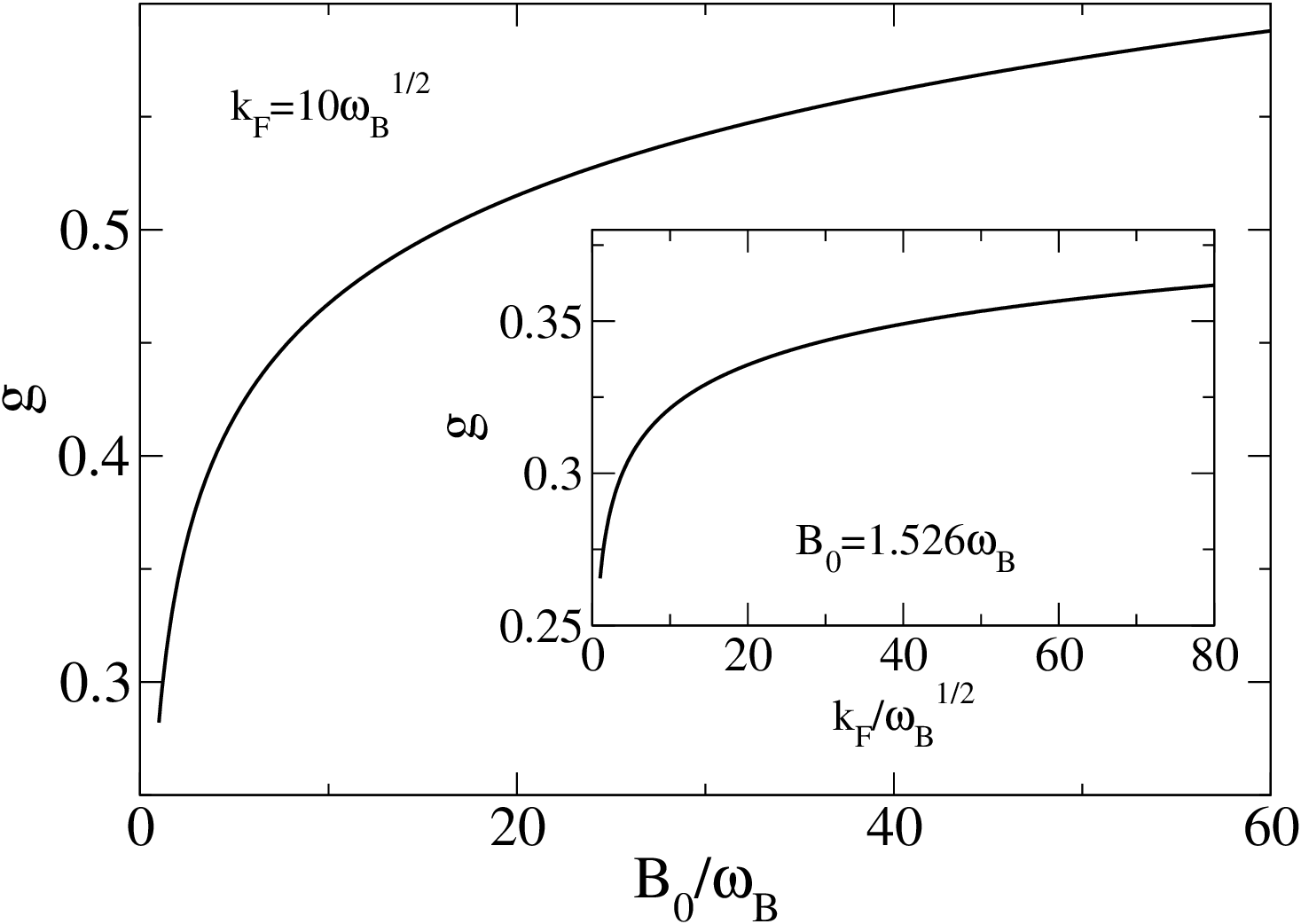}
\caption{\label{fig5}
TLL parameter $g$ in Eq.~(\ref{gsemicl}) for $\alpha=1$ and
$D=100/\sqrt{\omega_{\rm B}}$. Main panel: $g$ as a function of
$B_0/\omega_{\rm B}$ for $k_{\rm F}=10\sqrt{\omega_{\rm B}}$.
Inset: $g$ vs $k_{\rm F}/\sqrt{\omega_{\rm B}}$ for
$B_0=1.526\omega_{\rm B}$. Note that the regime $\lambda\ll d$
translates into $k_{\rm F}\gg\frac{18}{\sqrt{3}}\omega_{\rm
B}^2B_0^{-5/2}$, which here implies $k_{\rm F}/\sqrt{\omega_{\rm
B}} > 1$. }
\end{figure}

Combining Eqs.~(\ref{exponent}) to (\ref{calr}), we get
the analytical estimate for the TLL parameter,
\begin{equation}\label{gsemicl}
g\simeq \left[\frac{g_0-\ln(D/d)} {g_0+\ln(D/d)}\right]^{1/2}\;,
\end{equation}
in the regime $D\gg d\gg \lambda$, see Eq.~\eqref{regime}, where
$\lambda$ is given in Eq.~\eqref{ldef}. Here, we have
abbreviated
\begin{eqnarray} \label{g0def}
g_0&=&\frac{\pi}{2\alpha}+\sqrt{2}(c_1+2c_2)+4(\sqrt{2}c_1-1)
\ln\frac{D}{\lambda}\nonumber\\
&+&\ln\left(e^{C/2}\frac{D}{\lambda}\right)\;.
\end{eqnarray}
The TLL parameter \eqref{gsemicl} is depicted in
Fig.~\ref{fig5}, where $|\partial_k\varepsilon_{k_{\rm
F}}|=v_{\rm F}$ and a ``fine structure constant'' $\alpha=1$
have been assumed. However, according to Eq.~(\ref{g0def}),
changes in $\alpha$ may be compensated for via changes in
$\lambda$, i.e.\ by modification of $k_{\rm F}$ or of the
magnetic field parameters.

As shown in the inset of Fig.~\ref{fig5}, values of about
$g\approx 0.3$ are expected for the chosen value of $B_0$, with
no pronounced variations when changing $D/\lambda$. Such weak
sensitivity to $k_{\rm F}$ found in graphene is in stark
contrast to semiconducting wires, where $g$ can vary
significantly when changing the carrier density.\cite{whlkahm}
As discussed in the Introduction, similar values for $g$ as
compared to the values in Fig.~\ref{fig5} have been reported for
other TLL systems such as quantum wires\cite{yacoby} or
SWNTs.\cite{bockrath,swnt,swnt2} On the other hand, the main
panel in Fig.~\ref{fig5} demonstrates that $g$ can be widely
tuned in graphene wires by changing the snake-state separation
$2d$ in Eq.~\eqref{ddef} via the magnetic field parameters, in
particular by sweeping the background field $B_0$.

\section{Conclusions}\label{sec5}

In this paper, we have analyzed the effects of electron-electron
interactions on the electronic properties of magnetic waveguides
formed in suitable inhomogeneous magnetic fields in graphene.
When there are two parallel lines along which the magnetic field
vanishes, a pair of counterpropagating snake states can be
formed, which are ideal unidirectional (chiral) channels similar
to the edge states in quantum Hall bars. We have studied the
case of a smooth magnetic field profile across the wire, but
similar results are expected also for other profiles, e.g.\ for
piece-wise constant fields. Employing a combination of
analytical methods for the band structure and for the
eigenfunctions (which were checked against exact diagonalization
results) with a thermodynamical approach, we have obtained a
closed result for the non-universal TLL parameter, see
Eq.~\eqref{gsemicl} and Fig.~\ref{fig5}. This parameter then
determines the power-law exponents appearing in many observables
of interest, in particular in the energy-dependence of the
tunneling density of states. Quite remarkably, we have
uncovered that the snake orbits impose a periodic superstructure
that breaks Galilei invariance of the resulting wire, and
modifies the commonly used estimate for the TLL parameter
$g$. The finite ${\cal R}$-term in Eq.~\eqref{calr} reflects
this physics. We expect this correction to affect also
edge states in quantum Hall bars.
Typical values for $g$ found here are comparable
to the values reported for semiconductor quantum wires and
single-wall carbon nanotubes. We thus expect that non-Fermi
liquid behavior should be sufficiently pronounced to be
observable in systems based on inhomogeneous magnetic fields.

There are two main advantages regarding experimental
observability in our system when compared to previous TLL
realizations. First, the TLL parameter $g$ can be tuned over a
significant region by sweeping the strength of the homogeneous
part of the magnetic profile. This is illustrated in the main
panel of Fig.~\ref{fig5}. Second, the unavoidable presence of
disorder is not expected to affect the TLL behavior, since
right- and left-moving electrons are spatially separated. This
may allow, for the first time, to experimentally study the
physics of an ultraclean TLL state.

While there are similarities to the physics of quantum Hall edge
states, the TLL state discussed here is quite different from the
chiral Luttinger liquid discussed in the context of the
fractional quantum Hall effect.\cite{chang} The $g$ parameter in
the latter case is fixed by the bulk filling factor, while here
it is non-universal and tunable. From a conceptual point of
view, the quantum Hall situation is also more intricate because
of the coupling of edge states to bulk states.\cite{chang} Such
complications are absent for the TLL state discussed in this
paper. To conclude, we hope that our work motivates experiments
in this direction.

\acknowledgments

We thank T.\ Heinzel for discussions. WH would like to thank
H.\ Grabert for continuous support. This work has been supported
by the SFB -TR/12, by the ESF INSTANS program, and by the A.~v.~Humboldt
foundation.

\appendix
\section*{Numerical diagonalization}

In this Appendix, we provide some details on how the numerical
diagonalization mentioned in Sec.~\ref{sec2b} has been
implemented. From Eq.~(\ref{aix}), we first observe that exact
eigenfunctions $\phi(x)$ of Eq.~(\ref{schrodinger}) behave as
$\sim\exp\left[-\frac{x^{1+\nu}}{1+\nu}\right]$ for
$x\to\infty$. For example, the exact zero-energy state to
Eq.~\eqref{vnu3} is given as
$\:(e^{-\xi^4/4+b_0\xi^2/2+\kappa\xi},0)^T\:$ up to a
normalization factor.

However, it is clear that accurate eigenvalues $\varepsilon_k^2$
only require to maximize the overlap between the approximated
wavefunction and $\phi(x)$. For convenience, we thus use the
complete and orthonormal oscillator basis
\[
\varphi_n(\xi)=\frac{1}{\pi^{1/4}\sqrt{2^nn!}}
e^{-\xi^2/2}H_n(\xi)\;,
\]
where $H_n$ are Hermite polynomials.
Matrix elements of the operator $H^2$ in Eq.~(\ref{schrodinger}) in
this basis read ($b_0=B_0/\omega_B$)
\begin{eqnarray*}
&& \omega_{\rm B}^{-1} \langle\varphi_n|H^2|\varphi_{n'}\rangle=
(2n+1+\kappa^2)\delta_{nn'} \\ 
&& +\langle\varphi_n|\xi^{2\nu}-(1-b_0^2)\xi^2-2b_0\xi^{\nu+1}
|\varphi_{n'}\rangle \\
&& -\langle\varphi_n| 2k(\xi^{\nu}-b_0\xi)
-\tau_z(\nu\xi^{\nu-1}-b_0)|\varphi_{n'}\rangle \;,
\end{eqnarray*}
where we use that for integer $\gamma$ and even $n+n'+\gamma$ ($\Gamma$
denotes the Gamma function)
\begin{eqnarray*}
&& \langle\varphi_n|\xi^{\gamma}|\varphi_{n'}\rangle=
\sqrt{\frac{2^{n+n'}n\:!n'!}{\pi}}\:\sum_{m=0}^{[n/2]}\sum_{m'=0}^{[n'/2]}
\left(-\frac{1}{4}\right)^{m+m'} \\ &&\times\frac{\Gamma(\frac{1+n+n'+\gamma}
{2}-m-m')}{m!\:m'!\:(n-2m)!\:(n'-2m')!} \;,
\end{eqnarray*}
where $\langle\varphi_n|\xi^{\gamma}|\varphi_{n'}\rangle=0$
otherwise. The symbol $[n]$ denotes the largest integer smaller
or equal to $n$. Upon carrying out standard diagonalization for
$0\le n,n'\le 30$ basis functions, we obtain the energy
dispersion $\varepsilon_k$ in Figs.~\ref{fig1} and \ref{fig3}
for $\nu=2$ and $\nu=3$, respectively. To numerical accuracy,
all levels are found independent of $\tau_z=\pm 1$, although the
effective potentials in Eq.~(\ref{schrodinger}) differ
considerably. Only the zero-energy level $\varepsilon_k=0$ in
Fig.~\ref{fig3} (there is no zero-energy level in
Fig.~\ref{fig1}) belongs purely to the upper pseudo-spin
component $\tau_z=+1$ for the valley point $K$ chosen here.
With the numerical diagonalization, we also obtain the
eigenfunctions $\phi_k(x)$ and $\chi_k(x)$ of the Dirac
Hamiltonian (\ref{spinors}). In Figs.~\ref{fig2} and
\ref{fig4}, we show the resulting density profiles for $\nu=2$
and $\nu=3$, respectively. All figures nicely confirm the
overall picture developed in Sec.~\ref{sec2b}.


\begin{thebibliography}{10}

\bibitem{ferry05}
For reviews, see e.g.\ {\sl Quantum transport
 in ultrasmall devices}, NATO Adv.\ Studies Institute Series~B,
 Vol.~342, ed.\ by D.K. Ferry (Plenum, New York, 1995).

\bibitem{tubes}
{\sl Carbon nanotubes: Synthesis, Structure, Properties, and
Applications}, ed.\ by M.S. Dresselhaus, G. Dresselhaus,
and Ph. Avouris, Springer Topics in Applied Physics, Vol.~80
(Springer, Berlin, 2001).

\bibitem{solyom} J. S\'olyom, Adv.\ Phys.\ {\bf 28}, 201 (1979).

\bibitem{haldane} F.D.M. Haldane, J.\ Phys.\ C {\bf 14}, 2585 (1981).

\bibitem{voit} J. Voit, Rep.\ Prog.\ Phys.\ {\bf 57}, 977
 (1994).

\bibitem{gogolin} A.O. Gogolin, A.A. Nersesyan, and A.M. Tsvelik,
{\sl Bosonization and Strongly Correlated Systems} (Cambridge University
Press, 1998).

\bibitem{giamarchi} T. Giamarchi, {\sl Quantum Physics in One
 Dimension} (Oxford University Press, 2003).

\bibitem{yacoby} 
For very recent experimental reports on TLL behavior in semiconductor
quantum wires, see H. Steinberg, G. Barak, A. Yacoby, L.N. Pfeiffer,
K.W. West, B.I. Halperin, and K. Le Hur, Nature Physics {\bf 4}, 116 (2008),
and references therein.

\bibitem{bockrath} M. Bockrath, D.H. Cobden, J. Lu, A.G. Rinzler,
 R.E. Smalley, L. Balents, and P.L. McEuen, Nature {\bf 397}, 598
 (1999); Z. Yao, H.W.Ch. Postma, L. Balents, and C. Dekker,
{\sl ibid.} {\bf 402}, 273 (1999); B. Gao, A. Komnik, R. Egger,
D.C. Glattli, and A. Bachtold, Phys.\ Rev.\ Lett.\ {\bf 92}, 216804 (2004).

\bibitem{geim04} K.S. Novoselov, A.K. Geim, S.V. Morozov, D. Jiang,
 Y. Zhang, S.V. Dubonos, I.V. Grigorieva, and A.A. Firsov,
 Science {\bf 306}, 666 (2004);
 K.S. Novoselov, A.K. Geim, S.V. Morozov, D. Jiang, M.I. Katsnelson,
 I.V. Grigorieva, S.V. Dubonos, and A.A. Firsov, Nature {\bf 438},
 197 (2005); Y. Zhang, Y.-W. Tan, H.L. Stormer, and P. Kim,
 {\sl ibid.} {\bf 438}, 201 (2005); C. Berger, Z. Song, X. Li, X. Wu,
 N. Brown, C. Naud, D. Mayou, T. Li, J. Hass, A.N. Marchenkov,
 E.H. Conrad, P.N. First, and W.A. de Heer, Science {\bf 312},
 1191 (2006); F. Molitor, J. G\"uttinger, C. Stampfer, D. Graf,
 T. Ihn, and K. Ensslin, Phys.\ Rev.\ B {\bf 76}, 245426 (2007).

\bibitem{review}
For a review, see A.K. Geim and K.S. Novoselov, Nature
Materials {\bf 6}, 183 (2007).

\bibitem{Note} We absorb the factor $e/c$ into $A$ and $B$.

\bibitem{semenoff} G.W. Semenoff, Phys.\ Rev.\ Lett.\ {\bf 53}, 2449
 (1984); D.P. DiVincenzo and E.J. Mele, Phys.\ Rev.\ B {\bf 29}, 1685 (1984).

\bibitem{peeters93} F.M. Peeters and A. Matulis, Phys.\ Rev.\
 B {\bf 48}, 15166 (1993); N. Malkova, I. G\'omez, and
 F. Dom\'\i nguez-Adame, {\sl ibid.} {\bf 63}, 035317 (2001);
S.M. Badalyan and F.M. Peeters, {\sl ibid.} {\bf 64}, 155303 (2001);
J.M. Pereira, Jr., F.M. Peeters,
and P. Vasilopoulos, {\sl ibid.} {\bf 75}, 125433 (2007).

\bibitem{review2} S.J. Lee, S. Souma, G. Ihm, and K.J. Chang,
Physics Reports {\bf 394}, 1 (2004).

\bibitem{demartino} A. De Martino, L. Dell'Anna, and R. Egger,
 Phys.\ Rev.\ Lett.\ {\bf 98}, 066802 (2007).

\bibitem{peetersnew}
M. Ramezani Masir, P. Vasilopoulos, A. Matulis, and F.M. Peeters,
Phys.\ Rev.\ B {\bf 77}, 235443 (2008). 

\bibitem{muller} J.E. M\"uller, Phys.\ Rev.\ Lett.\ {\bf 68}, 385 (1992);
J. Reijiniers and F.M. Peeters, J. Phys.: Condens.\ Matter
{\bf 12}, 9771 (2000); J. Reijiniers, A. Matulis, K. Chang,
F.M. Peeters, and P. Vasilopoulos, Europhys.\ Lett.\ {\bf 59}, 749 (2002);
H. Xu, T. Heinzel, M. Evaldsson, S. Ihnatsenka,
and I.V. Zozoulenko, Phys.\ Rev.\ B {\bf 75}, 205301 (2007).

\bibitem{singlesnake} P.D. Ye, D. Weiss, R.R. Gerhardts, M. Seeger,
 K. von Klitzing, K. Eberl, and H. Nickel, Phys.\ Rev.\ Lett.\
 {\bf 74}, 3013 (1995); D. Lawton, A. Nogaret, M.V. Makarenko, O.V.
 Kibis, S.J. Bending, and M. Henini, Physica~E, {\bf 13}, 699 (2002);
M. Hara, A. Endo, S. Katsumoto, and Y. Iye, Phys.\ Rev.\ B {\bf 69}, 
153304 (2004); M. Cerchez, S. Hugger, T. Heinzel, and N. Schulz,
{\sl ibid.} {\bf 75}, 035341 (2007).

\bibitem{lambert} L. Oroszl\'any, P. Rakyta, A. Korm\'anyos,
 C.J. Lambert, and J. Cserti, Phys.\ Rev.\ B {\bf 77}, 081403(R) (2008).

\bibitem{ghosh} T.K. Ghosh, A. De Martino, W. H\"ausler, L. Dell'Anna,
 and R. Egger, Phys.\ Rev.\ B {\bf 77}, 081404(R) (2008).

\bibitem{cserti08}
A. Korm\'anyos, P. Rakyta, L. Oroszl\'any, and J. Cserti,
arXiv:0805.2527.

\bibitem{snaketube}
H.-W. Lee and D.S. Novikov, Phys.\ Rev.\ B {\bf 68}, 155402 (2003);
N. Nemec and G. Cuniberti, {\sl ibid.} {\bf 74}, 165411 (2006).

\bibitem{footnew}
In thermal equilibrium, unidirectional 1D states must occur in pairs 
of counterpropagating snake (or edge) modes.\cite{lambert,ghosh}

\bibitem{doublesnake} A. Nogaret, S.J. Bending, and M. Henini,
 Phys.\ Rev.\ Lett.\ {\bf 84}, 2231 (2000).

\bibitem{bostwick} A. Bostwick, T. Ohta, T. Seyller, K. Horn, and
E. Rotenberg, Nature Physics {\bf 3}, 36 (2007).

\bibitem{FQHE} M.O. Goerbig, R. Moessner, and B. Dou\c{c}ot,
 Phys.\ Rev.\ B {\bf 74}, 161407(R) (2006);
 K. Nomura and A.H. MacDonald, Phys.\ Rev.\ Lett.\
 {\bf 96}, 256602 (2006); L. Sheng, D.N. Sheng, F.D.M. Haldane,
and L. Balents, {\sl ibid.} {\bf 99}, 196802 (2007);
C.-H. Zhang and Y.N. Joglekar, Phys.\ Rev.\ B {\bf 75}, 245414 (2007).

\bibitem{falko06} 
T. Ando, J. Phys. Soc. Jpn. {\bf 75}, 074716 (2006); 
V.V. Cheianov and V.I. Fal'ko, Phys.\ Rev.\ Lett.\ {\bf 97}, 226801 (2006);
E.G. Mishchenko, {\sl ibid.} {\bf 98}, 216801 (2007);
Y. Barlas, T. Pereg-Barnea, M. Polini, R. Asgari, and A.H. MacDonald,
{\sl ibid.} {\bf 98}, 236601 (2007);
E. Mariani, L.I. Glazman, A. Kamenev, and
F. von Oppen, Phys.\ Rev.\ B {\bf 76}, 165402 (2007).

\bibitem{vozmediano} J. Gonzalez, F. Guinea, and M.A.H. Vozmediano,
Nucl.\ Phys.\ B {\bf 424}, 595 (1994); Phys.\ Rev.\ B {\bf 63}, 134421 (2001);
T. Stauber, F. Guinea, and M.A.H. Vozmediano, {\sl ibid.} {\bf 71},
041406(R) (2005); D.T. Son, {\sl ibid.} {\bf 75}, 235423 (2007).

\bibitem{khveshchenko06}
I.F. Herbut, Phys.\ Rev.\ Lett.\ {\bf 97}, 146401 (2006);
D.V. Khveshchenko, Phys.\ Rev.\ B {\bf 74}, 161402(R) (2006);
S. Das Sarma, E.H. Hwang, W.-K. Tse, {\sl ibid.} {\bf 75}, 121406 (2007);
I.F. Herbut, V. Juricic, and O. Vafek,
Phys.\ Rev.\ Lett.\ {\bf 100}, 046403 (2008).

\bibitem{ribbons}
H.A. Fertig and L. Brey, Phys.\ Rev.\ Lett.\ {\bf 97}, 116805 (2006);
M. Zarea and N. Sandler, {\sl ibid.} {\bf 99}, 256804 (2007).

\bibitem{swnt} R. Egger and A.O. Gogolin, Phys.\ Rev.\ Lett.\
 {\bf 79}, 5082 (1997); C. Kane, L. Balents, and M.P.A. Fisher,
 {\sl ibid.} {\bf 79}, 5086 (1997).

\bibitem{swnt2} R. Egger and A.O. Gogolin,
 Eur.\ Phys.\ J.\ B {\bf 3}, 281 (1998).

\bibitem{safi} I. Safi and H.J. Schulz,
 Phys.\ Rev.\ B {\bf 52}, R17040 (1995).

\bibitem{noise} B. Trauzettel, R. Egger, and H. Grabert, Phys.\ Rev.\ Lett.\
 {\bf 88}, 116401 (2002); F. Dolcini, B. Trauzettel, I. Safi, and
 H. Grabert, Phys.\ Rev.\ B {\bf 71}, 165309 (2005).

\bibitem{hsquared}
 L.C. Biedenharn, Phys.\ Rev.\ {\bf 126}, 845 (1962);
 G.J. Papadopoulos and J.T. Devreese, Phys.\ Rev.\ D {\bf 13}, 2227 (1976);
 M.A. Kayed and A. Inomata, Phys.\ Rev.\ Lett.\ {\bf 53}, 107 (1984);
 T. Boudjedaa, L. Chetouani, L. Guechi, and T.F. Hammann,
 Physica Scripta {\bf 46}, 289 (1992).

\bibitem{ullmo}
P. Carmier and D. Ullmo, arXiv:0801.4727.

\bibitem{indexthyrefs} Y. Aharonov and A. Casher, Phys.\ Rev.\ A {\bf
 19}, 2461 (1979); L. Erd\"os and V. Vougalter, Commun.\ Math.\
 Phys.\ {\bf 225}, 299 (2002).

\bibitem{park08} S. Park and H.-S. Sim, Phys.\ Rev.\ B {\bf 77},
 075433 (2008).

\bibitem{larkin74} I.E. Dzyaloshinski\u{\i} and A.I. Larkin,
 Zh.\ \'Eksp.\ Teor.\ Phys.\ {\bf 65}, 411 (1973) [Sov.\ Phys.\
 JETP {\bf 38}, 202 (1974)].

\bibitem{qheref} B.I. Halperin, Phys.\ Rev.\ B {\bf 25}, 2185 (1982);
 A.H. MacDonald, Phys.\ Rev.\ Lett.\ {\bf 64}, 220 (1990).

\bibitem{chang} A.M. Chang, Rev.\ Mod.\ Phys.\ {\bf 75}, 1449 (2003).

\bibitem{UZ} U. Z\"ulicke and A.H. MacDonald,
 Phys.\ Rev.\ B {\bf 54}, 16813 (1996).

\bibitem{whlkahm} W. H\"ausler, L. Kecke, and A.H. MacDonald,
 Phys.\ Rev.\ B {\bf 65}, 085104 (2002).

\bibitem{alicea} J. Alicea and M.P.A. Fisher, Phys.\ Rev.\ B {\bf 74},
 075422 (2006).

\bibitem{noWC} H.P. Dahal, Y.N. Joglekar, K.S. Bedell, and
 A.V. Balatsky, Phys.\ Rev.\ B {\bf 74}, 233405 (2006).

\bibitem{epl} C.E. Creffield, W. H\"ausler, and A.H. MacDonald,
 Europhys.\ Lett.\ {\bf 53}, 221 (2001).

\bibitem{ernst97} G. Ernst, N.B. Zhitenev, R.J. Haug, and
 K. von Klitzing, Phys.\ Rev.\ Lett.\ {\bf 79}, 3748 (1997).

\bibitem{wh} This also applies to multiband situations, cf.\
 W. H\"ausler, Phys.\ Rev.\ B {\bf 70}, 115313 (2004).

\bibitem{liebwu} E. Lieb and F.Y. Wu, Phys.\ Rev.\ Lett.\
 {\bf 20}, 1445 (1968).

\bibitem{kawakami91} N. Kawakami and S.-K. Yang, Phys.\ Rev.\ Lett.\
 {\bf 67}, 2493 (1991).

\bibitem{lkwh} L. Kecke and W. H\"ausler,
 Phys.\ Rev.\ B {\bf 69}, 085103 (2004).

\end{thebibliography}
\end{document}